\UseRawInputEncoding
\documentclass[reprint, superscriptaddress, prl]{revtex4-2}
\usepackage{lineno}
\usepackage[separate-uncertainty=true,multi-part-units=single]{siunitx}
\usepackage{graphicx} % Include figure files
\usepackage{amsmath}
\usepackage{xcolor} % colors for comments
\usepackage{hyperref} % add links
\DeclareSIUnit{\dBm}{dBm}

\newcommand{\PME}{\affiliation{Pritzker School of Molecular Engineering, University of Chicago, Chicago IL 60637, USA}}
\newcommand{\ucphysics}{\affiliation{Department of Physics, University of Chicago, Chicago IL 60637, USA}}

\begin{document}
	%\linenumbers
	
	\title{Bidirectional multi-photon communication between remote superconducting nodes}

	\author{Joel Grebel}\PME
 	\author{Haoxiong Yan}\PME
	\author{Ming-Han Chou}\PME \ucphysics
	\author{Gustav Andersson}\PME
	\author{Christopher R. Conner}\PME
	\author{Yash J. Joshi}\PME
	\author{Jacob M. Miller}\PME \ucphysics
	\author{Rhys G. Povey}\PME \ucphysics
	\author{Hong Qiao}\PME
	\author{Xuntao Wu}\PME
	\author{Andrew N. Cleland}\PME
	\email{anc@uchicago.edu}
	\affiliation{Center for Molecular Engineering, Argonne National Laboratory, Lemont IL 60439, USA}

\date{\today}
	
\begin{abstract}
	Quantum communication testbeds provide a useful resource for experimentally investigating a variety of communication protocols. Here we demonstrate a superconducting circuit testbed with bidirectional multi-photon state transfer capability using time-domain shaped wavepackets. The system we use to achieve this comprises two remote nodes, each including a tunable superconducting transmon qubit and a tunable microwave-frequency resonator, linked by a 2 m-long superconducting coplanar waveguide, which serves as a transmission line. We transfer both individual and superposition Fock states between the two remote nodes, and additionally show that this bidirectional state transfer can be done simultaneously, as well as used to entangle elements in the two nodes. 
\end{abstract}
	
\maketitle

%%%%% INTRODUCTION
Long range, high fidelity communication of quantum information has applications in several areas \cite{Gisin2007}, including secure communication using quantum key cryptography \cite{Ekert1991}, as well as serving as the backbone for a future quantum internet \cite{Kimble2008}. These applications require sources of entangled photons, preferably on-demand, to perform most quantum cryptographic functions. Photons at optical frequencies are a natural choice for the communication medium, due to their high energies compared to ambient thermal energies, low propagation loss at room temperature, and widely available fiber communication technology.  However, high fidelity and high rate sources of on-demand entangled photons are still lacking at optical frequencies, yielding to date low information transfer rates \cite{Reiserer2022, Briegel1998, Munro2015, Yin2017, Du2021}. At microwave frequencies, superconducting circuits provide a flexible platform for designing high fidelity control of computational elements with reasonably low-loss memory elements, and can deterministically generate microwave photons entangled with qubits. Combined with variable superconducting couplers, these elements can be used to experimentally test long-range communication protocols \cite{Casariego2023} with itinerant photon wavepackets \cite{Yin2013,Wenner2014, Pierre2014,Srinivasan2014, Kurpiers2018, CampagneIbarcq2018, Zhong2019, Axline2018, Magnard2020, Qiu2023}, as well as modular quantum computing approaches, using the standing modes in a weakly-coupled communication waveguide \cite{Chang2020, Burkhart2021, Zhong2021, Niu2023}. Several experiments using superconducting circuits have demonstrated deterministic transfer of both single \cite{Kurpiers2018, CampagneIbarcq2018, Zhong2019} and multi-photon states \cite{Axline2018} between remote nodes, using time-symmetric shaped wavepackets to improve transfer fidelity \cite{Cirac1997}. However, with the exception of Ref. \onlinecite{Zhong2019} and \onlinecite{Qiu2023}, these all involved the use of lossy microwave circulators, thus only supporting communication in one direction.
	
Here we demonstrate bidirectional, multi-photon state transfer with shaped wavepackets between two remote superconducting qubit nodes, eliminating the microwave circulators used in earlier experiments \cite{Kurpiers2018,CampagneIbarcq2018,Axline2018}. Circulators are useful for preventing unwanted reflections from emitted signals, simplifying tune-up and operation of these circuits. However, existing broadband commercial circulators cannot reverse their polarity \emph{in situ}, and further are a significant source of loss; while broadband parametric circulators may overcome these limitations \cite{Beck2022,Kwende2023}, these are not yet available. Here we implement itinerant bidirectional communication, where itinerant means the pulsed signals have time-domain envelopes shorter than the length of the transmission line; we achieve this by using fast dynamic couplers at each node, allowing us to complete signal transfers with low reflection rates  and sufficient speed that interference between emission and reflection signals can be avoided \cite{Zhong2019}.  
	
\begin{figure}
	\centering
	\includegraphics[width=3.37in]{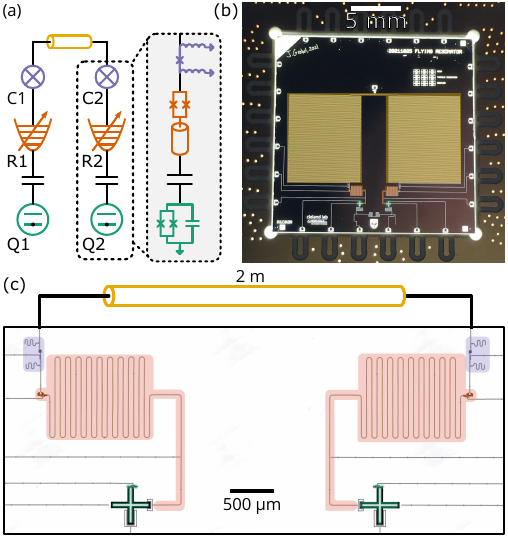}
	\caption{Experimental layout. (a) Block diagram for the experiment, comprising two tunable qubits $Q1$, $Q2$ (green), capacitively-coupled to two tunable resonators $R1$, $R2$ (orange), in turned coupled via two variable couplers $C1$, $C2$ (purple) to a \qty{2}{m}-long waveguide (yellow). Circuit representations of these elements are shown in the dashed box. (b) Backside-illuminated optical photograph of fabricated device with false coloring, showing all components including the \qty{2}{m}-long coplanar waveguide, as well as the printed circuit board to which the die is wirebonded for signal routing. (c) Optical micrograph of device with false coloring corresponding to components shown in panel (a). The coplanar waveguide is patterned on the same die but is cropped from this image; see panel (b).}
	\label{fig:sample}
\end{figure}

The experimental device is shown in Fig.~\ref{fig:sample}. Each of the two nodes comprises a frequency-tunable superconducting Xmon qubit \cite{Barends2013} $Q1$ ($Q2$), capacitively-coupled to a frequency-tunable resonator $R1$ ($R2$), which is in turn connected via a variable coupler \cite{Chen2014} to one end of a 2 m-long, $50~\Omega$ coplanar waveguide. The tunable resonators are implemented as a resonant section of coplanar waveguide terminated by a superconducting quantum interference device (SQUID) \cite{Sandberg2008}, connected to ground through the variable coupler.  Control flux lines allow tuning the resonator over a range of \qty{1.5}{GHz}, limited in the experiment to 0.5 GHz by the control electronics. The variable couplers afford control of the coupling strength of the resonator to the 2-m long transmission line, where the resonator decay rate $\kappa_r/2\pi$ into the transmission line can be varied dynamically from $0$-$\qty{55}{MHz}$, with control signals as short as \qty{3}{ns}, limited by control-line filtering. The transmission line itself is a $50~\Omega$ coplanar waveguide, galvanically connected to each variable coupler. All components are fabricated on a single \qty{2}{cm} $\times$ \qty{2}{cm} sapphire die, shown in a back-lit optical micrograph in Fig.~\ref{fig:sample}(b). Fabrication details are provided in Ref.~\onlinecite{Supplement}. The device was wirebonded to a copper printed circuit board that was in turn placed in an aluminum enclosure, the latter placed in a double magnetic shield mounted to the \qty{10}{mK} mixing chamber of a dilution refrigerator. A wiring diagram showing the cabling for the dilution refrigerator, and a schematic for the control and readout circuitry, is shown in the supplemental material \cite{Supplement}.
	
\begin{figure}
	\centering
	\includegraphics[width=3.37in]{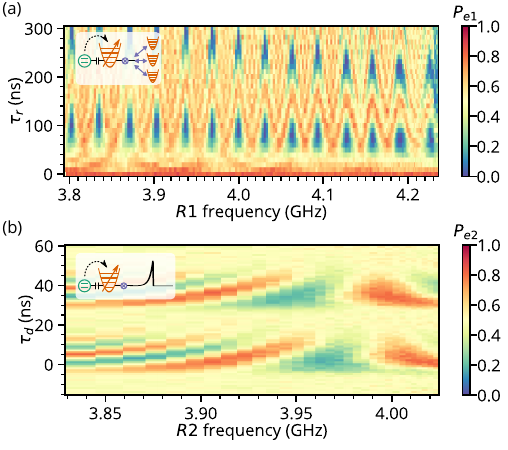}
	\caption{  {(a)} Standing waveguide modes seen by $R1$ at weak coupling. After a single excitation is swapped into $R1$, the coupling to the waveguide is turned on for time $\tau_r$ while tuning the resonator frequency. {(b)} Ramsey-like experiment showing itinerant emission and recapture of a $|0\rangle + |1\rangle$ state in $R2$. The superposition state is swapped from $Q2$ to $R2$, emitted and recaptured from $R2$ with \qty{20}{ns} rectangular pulses separated by delay $\tau_d$ while tuning the resonator frequency, swapped back to $Q2$, and a $\pi/2$ pulse is applied to $Q2$ before measurement.}
	\label{fig:characterization}
\end{figure}

Standard characterization of the qubits yields characteristic lifetimes $ T_{1,Q1} = \qty{20}{\mu s}$ and $ T_{1,Q2} = \qty{22}{\mu s} $ at the qubit operating frequencies of $f_{Q1} = \qty{4.57}{GHz}$ and $f_{Q2}=\qty{4.5}{GHz}$. The qubit Ramsey lifetimes $T_2$ at the same operating frequencies are \qty{2.6}{\mu s} and \qty{0.56}{\mu s}, respectively. 

To characterize the tunable resonators, we swap single excitations from the qubits to their respective resonators by tuning the excited qubit into resonance with the resonator for a calibrated time, with the variable couplers turned off. Using the qubits to monitor the subsequent resonator decay, we measure characteristic $T_1$ and $T_2$ times of $T_{1,R1} = \qty{4.57}{\mu s},$ $T_{1,R2} = \qty{0.86}{\mu s},$ $ T_{2,R1} = \qty{0.95}{\mu s}, $ and $T_{2,R2} = \qty{0.9}{\mu s}$. All qubit-resonator swaps are measured at frequency $f_R = \qty{4.058}{GHz}$ with qubit-resonator coupling strength $g_{QR}/2\pi = \qty{6.8}{MHz}$ set by the geometric capacitance C = \qty{1.4}{fF} connecting these elements. 

We characterize the transmission waveguide by swapping excitations into individual standing waveguide modes as shown in Fig.~\ref{fig:characterization}(a), at weak coupling $g_{RW} \ll \omega_{FSR}$, with $g_{RW}/2\pi < \qty{3.4}{MHz}$ the resonator/waveguide coupling strength, and $\omega_{FSR}/2\pi = \qty{31}{MHz}$ the waveguide free spectral range. The waveguide $T_1$ coherence times are in the range of $4-\qty{5}{\mu s}$. At stronger coupling, we emit itinerant wavepackets into the waveguide, which requires coupling to a number of adjacent waveguide modes. Figure \ref{fig:characterization}(b) shows a Ramsey-like experiment with $\pi/2$ pulses on $Q2$ at the beginning and end of the pulse sequence, where  we emit and then recapture a superposition $|0\rangle + |1\rangle$ state in $R2$ for different resonator frequencies. The return time for the pulses is independent of the resonator frequency in this regime; fringes at different frequencies indicate phase coherence in the waveguide.  See Ref.~\onlinecite{Supplement} for more characterization details. 
 
\begin{figure}
	\centering
	\includegraphics[width=3.37in]{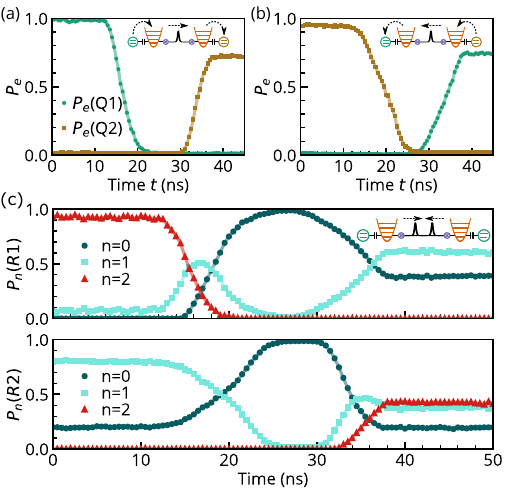}
	\caption{Transfer of excited qubit states, from (a) qubit $ Q1 $ to qubit $ Q2 $ and (b) $ Q2 $ to $ Q1 $. We excite the emitting qubit, swap the excitation to the adjacent resonator, transmit the excitation as a wavepacket to the distant resonator by time-varying control of the couplers between the resonators and waveguide, then finally swap to the receiving qubit. Lines are a guide for the eye. (c) Simultaneous state swaps between the resonators. We initialize resonator $R1$ in a 2-photon Fock state while initializing resonator $R2$ in a 1-photon Fock state. These states are simultaneously released into the transmission line as itinerant wavepackets that are then captured by the receiving resonators, and we use the associated qubit to analyze its associated resonator's final state. The population of each resonator's Fock states $|F_n\rangle$, $n = 0, 1, 2$ is shown as a function of time. }
	\label{fig:bidirectional}
\end{figure}
	
We prepare Fock states in each tunable resonator by exciting the adjacent qubit, then resonantly swapping photons one at a time into the resonator \cite{Hofheinz2008}. In Fig.~\ref{fig:bidirectional}(a) and (b), we swap an $n=1$ Fock state between the resonators via the communication waveguide, by first exciting either resonator via its adjacent qubit with both resonators tuned to $f_R$, and controlling the variable couplers with calibrated pulses to perform an itinerant release-and-catch transfer of the excitation \cite{Zhong2019}. We then swap the receiving resonator's excitation to the corresponding qubit for measurement. We release and subsequently capture the wavepacket by dynamically tuning the coupling strength for each coupler, and optimize the state transfer using Bayesian optimization \cite{Nogueira2014}. During this process, we hold each resonator's frequency constant by applying a stabilizing flux pulse to its SQUID; this corrects both for frequency changes due to the changing loading electrical circuit as well as flux cross-talk. We transfer single photons in either direction, with an efficiency of 0.72 for $Q1 \rightarrow Q2$ transfers, and an efficiency of 0.74 for the reverse direction. Here efficiency is defined as $\mathcal{E} = \langle P_j \rangle / \langle P_i \rangle$, where $P_{i,j}$ are the release and capture qubit populations, respectively. The inefficiency is likely dominated by non-ideal pulse shaping and timing, as the inverse loss rate in the system is well-characterized and significantly longer than the transfer time.
 
To demonstrate bidirectional itinerant state transfer, we prepare dual resonator states using the adjacent qubits and then simultaneously release these into the communication waveguide, using the qubits to monitor the resonator populations. In Fig.~\ref{fig:bidirectional}(c) we show the schematic process, where we prepare resonator $R1$ with a two-photon Fock state while preparing resonator $R2$ with a one-photon Fock state. We then use both variable couplers to perform a simultaneous itinerant release-and-catch of photons transmitted in each direction in the communication waveguide.  To reconstruct the resulting population in the resonators, we resonantly interact the resonator with its corresponding qubit for a variable length of time, monitoring the qubit excited state probability. We then fit the time-dependent response to a model Hamiltonian for the combined system \cite{Hofheinz2009}. The reconstructed resonator Fock state probabilities for Fock states $n=0, 1, 2$ are shown in Fig.~\ref{fig:bidirectional} as a function of time. Due to the use of slightly different control pulses, in this experiment the resonator one-photon transfer efficiency is 0.82, while the two-photon transfer efficiency is 0.64.
	
\begin{figure}
	\centering
	\includegraphics[width=3.37in]{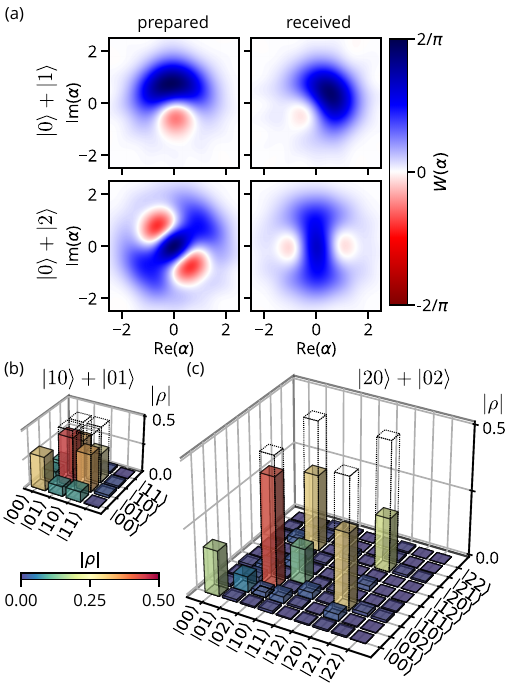}
	\caption{(a) Transfer of superposition Fock states. We prepare $ |0\rangle + |1\rangle$ and $ |0\rangle + |2\rangle$ states in resonator $R1$ and transfer these as itinerant wavepackets to resonator $R2$. We use the corresponding qubits to reconstruct the Wigner tomograms in both resonators, as described in the main text.  NOON states $|n0\rangle + |0n\rangle$ with  (b) $n=1$ and (c) $n=2$ itinerant transfers between the two tunable resonators. State preparation is described in the main text. Dotted lines indicate the ideal NOON states while transparent colored bars are the reconstructed density matrices.}
	\label{fig:superpositionentangle}
\end{figure}

We can also use this system to generate and transfer more complex quantum states. In Fig.~\ref{fig:superpositionentangle}(a), we show the preparation and transfer of superposition Fock states. We prepare the $ |0\rangle + |1\rangle $ superposition states\ in resonator $R1$ by swapping a $ |g\rangle + |e\rangle $ state from qubit $Q1$, achieving a state fidelity  $\mathcal{F} = 0.99(6)$. We perform an itinerant state transfer via the communication channel to resonator $R2$ with fidelity $\mathcal{F} = 0.94(0)$, where $\mathcal{F}(\rho, \sigma) = \text{tr}\sqrt{\rho^{1/2} \sigma \rho^{1/2}} $ of the measured density matrix $\rho$ to the ideal state $\sigma$.  In a similar fashion, we prepare the $ |0\rangle + |2\rangle $ superposition in $R2$ by exciting $Q1$ to the $ |g\rangle + |f\rangle $ state with two subsequent swaps to $R1$, achieving a fidelity $\mathcal{F} = 0.95(2)$. The $ |0\rangle + |2\rangle $ superposition is then transferred by wavepacket via the communication channel to $R1$, with fidelity $\mathcal{F} = 0.82(3)$. The Wigner tomograms in the figure are reconstructed by using convex optimization to find the most likely state that fits the Fock distribution as a function of displacement \cite{Grant2014,Strandberg2022}.  To measure the final resonator Fock state distribution, we tune the qubit, initially in its ground state, to the resonator frequency, and fit the subsequent qubit oscillations to a model system \cite{Hofheinz2009,Supplement}. The phase difference between the prepared and received states are possibly due to the resonator frequency varying slightly during wavepacket release and capture, due to non-ideal control pulses. Wigner tomograms of the raw parity data are given in \cite{Supplement}.
	
To demonstrate the generation of remote entanglement in this system, shown in Fig.~\ref{fig:superpositionentangle} (b), we create NOON states, superposition states in which $ N $-photon states in one resonator are superposed with vacuum states in the other resonator, in the form $ |\psi\rangle = \frac{1}{\sqrt{2}} (|N0\rangle - |0N\rangle) $. In Fig.~\ref{fig:superpositionentangle} (b), for the $N=1$ NOON state, we prepare a one-photon Fock state in $R1$, then, via an itinerant transfer, move half the population to $R2$ . We then measure the joint resonator state with bipartite Wigner tomography \cite{Wang2011}, using convex optimization to find the most likely state given by the joint resonator Fock distribution as a function of displaced resonator states \cite{Grant2014,Strandberg2022}. Similar to the single resonator case, we measure the joint resonator Fock distribution by tuning both qubits, initially in their ground states, into resonance with their respective resonators, then fit the resulting qubit population oscillations to a model system \cite{Wang2011, Supplement}. The fidelity of the prepared state is found to be $\mathcal{F} = 0.82(7)$. In Fig.~\ref{fig:superpositionentangle} (c), for the $ N=2 $ NOON state, we first prepare a one-excitation Bell state $(|eg\rangle-|ge\rangle)/\sqrt{2}$ in the two qubits, by swapping half an excitation from $Q1$ to $R1$, followed by an itinerant transfer from $R1$ to $R2$, and finally swapping the half-excitation to $Q2$. We next excite the $ e-f $ transition in each qubit, resulting in the state $(|fg\rangle-|gf\rangle)/\sqrt{2}$, then transfer the qubit populations to the resonators with two subsequent swaps, resulting in the final resonator state $(|20\rangle-|02\rangle)/\sqrt{2}$. The fidelity to the ideal state is $\mathcal{F} = 0.78(0)$. Fig.~\ref{fig:superpositionentangle} (b) and (c) show the reconstructed absolute value of the density matrices for the two states. The major sources of infidelity are photon loss during the transfer process and while stored in the resonators. 
	
%%%%% CONCLUSION
In conclusion, we demonstrate multi-photon bidirectional communication between two quantum-coherent superconducting nodes coupled by a 2 m-long coplanar waveguide, with states sent using itinerant photons whose pulse lengths of \qty{0.45}{m} are significantly less than the 2 m length of the waveguide. Future experiments might transfer multi-photon qubits such as the cat \cite{Lescanne2020} or GKP \cite{CampagneIbarcq2020} encoding states between resonators in the testbed.  Other communication protocols might be achieved by dynamically varying both the resonator frequencies and the coupling strength of each resonator to the waveguide during wavepacket emission and capture.  We can judge the relative performance of different communication protocols using a single system, e.g. using final state fidelities. Tens of individual waveguide modes can be addressed by each node; this system thus has further potential as a quantum random access memory \cite{Hann2019}. Additional nodes may be connected to the network by adding multiple coupled waveguides to each resonator, with nodes in separate modules if needed \cite{Zhong2021}.  The most significant limitations in this system are the challenges in the time-domain control of both the couplers and resonators, which has to account for circuit loading, flux cross-talk and cable-related pulse distortions. Higher fidelities might be achieved using machine learning techniques such as implemented in Ref. \cite{Sivak2022,Wu}; more precise pulse control with faster electronics and fewer control wiring filters; and a longer waveguide that allows longer duration itinerant photon pulse shapes, making time-domain control less challenging.
	
	%%%%% END MATTER
\section{Acknowledgments}
We thank P. J. Duda for helpful discussions and W. D. Oliver and G. Calusine at Lincoln Laboratories for the provision of a traveling-wave
parametric amplifier. Financial support was provided by the NSF QLCI for HQAN (NSF Award 2016136), the U.S. Department of Energy Office of Science National Quantum Information Science Research Centers, the Army Research Office-Laboratory for Physical Sciences (contract W911NF-23-1-0077), and the University of Chicago MRSEC (NSF award DMR-2011854). We made use of the Pritzker Nanofabrication Facility, partially supported by SHyNE, a node of the National Science Foundation’s National Nanotechnology Coordinated Infrastructure (NSF Grant No. NNCI ECCS-2025633). A.N.C. was supported in part by the DOE, Office of Basic Energy Sciences. The authors declare no competing financial interests. Correspondence and requests for materials should be addressed to A. N. Cleland (anc@uchicago.edu).

J.G. designed and fabricated the devices, performed the experiment and analyzed the data. H.Y., M.H.C., and G.A. provided suggestions for measurement and data analysis. A.N.C. advised on all efforts. All authors contributed to discussion and production of the manuscript.

\bibliography{references}
	
\end{document}

\title{Supplemental Material for ``Bidirectional multi-photon communication between remote superconducting nodes"}

\author{J. Grebel}\PME
\author{G. Andersson}\PME
\author{M.-H. Chou}\PME \ucphysics
\author{C. R. Conner}\PME
\author{Y. J. Joshi}\PME
\author{J. M. Miller}\PME \ucphysics
\author{R. G. Povey}\PME \ucphysics
\author{H. Qiao}\PME
\author{X. Wu}\PME
\author{H. Yan}\PME
\author{A. N. Cleland}\PME
\email{anc@uchicago.edu}
\affiliation{Center for Molecular Engineering, Argonne National Laboratory, Lemont IL 60439, USA}

\date{\today}
\maketitle
\onecolumngrid
\section{Device fabrication and wiring}

\subsection{Fabrication}
After depositing \qty{100}{nm} of Al by electron beam evaporation on a sapphire substrate \cite{Kelly2015,Dunsworth2017}, the qubits, resonators, and waveguide are patterned with photolithography and etched with a BCl$_3$/Cl$_2$/Ar inductively coupled plasma (ICP). We locate any unwanted waveguide shorts both electrically with an on-chip probe and visually with a microscope. We then etch away these shorts with a targeted second photolithography process of the base layer and Cl etch. For the airbridge sacrificial layer, \qty{200}{nm} of SiO$_2$ is electron beam evaporated on patterned photoresist followed by a N-methyl-2-pyrrolidone (NMP) liftoff at $80\ ^\circ$C. We deposit Josephson junctions with a Dolan bridge process using a PMMA/MAA bilayer patterned with electron beam lithography, aligned on the wafer using photolithographically defined markers (\qty{10}{nm} Ti, \qty{150}{nm} Au). A final photolithography process defines both Josephson junction bandages and the top airbridge layer; following which the sample is Ar-ion milled and \qty{200}{nm} of Al deposited with electron beam evaporation. After dicing individual device dies, we remove the sacrificial SiO$_2$ with vapor HF to form suspended airbridges \cite{Dunsworth2018}.  More detailed fabrication information can be found in Ref.~\cite{Grebel2023}.

\subsection{Resonant package modes}
All components are fabricated on a single $2 \times 2$ \qty{}{cm} sapphire substrate. The unusually large die size is needed to fit the entire transmission line. Unfortunately, at this die size, the box modes of both the die and package are near experimentally relevant frequencies with the frequencies $ f_{nml} $ given by \cite{Huang2021}
\begin{equation}
	f_{nml} = \frac{c}{2 \pi \sqrt{\mu_r \epsilon_r}}\sqrt{\left(\frac{n \pi}{a}\right)^2 + \left(\frac{m \pi}{b}\right)^2 + \left(\frac{l \pi}{d}\right)^2} , 
\end{equation}
where $ n,m, $ and $ l $ are the mode numbers for the transverse electric (TE) and transverse magnetic (TM) modes in a rectangular cavity, with two out of three indices non-zero. The relative permeability and permittivity are given as $ \mu_r $ and $ \epsilon_r $, $ c $ is the speed of light and $ a,b, $ and $ d $ are the cavity dimensions (for example along the Cartesian $x,y,z$ directions).

For a $ \qty{20}{mm} $ square sapphire die with $ \epsilon_r = 11.4 $, the lowest frequency mode is $f_{110, chip} =\qty{3.14}{GHz}$, while using $ \epsilon_r=1 $ and $ \qty{27}{mm}$ square for the package mode gives $ f_{110 package} =\qty{7.85}{GHz}$. From this rough calculation, the die modes are within our typical superconducting circuit operating frequencies of 4-\qty{8}{GHz} while the package modes are above our operating frequencies. We tune the frequencies of our qubits and resonators to avoid these modes.

\subsection{Wiring}
\begin{figure*}[ht]
	\centering
	\includegraphics[width=6.74in]{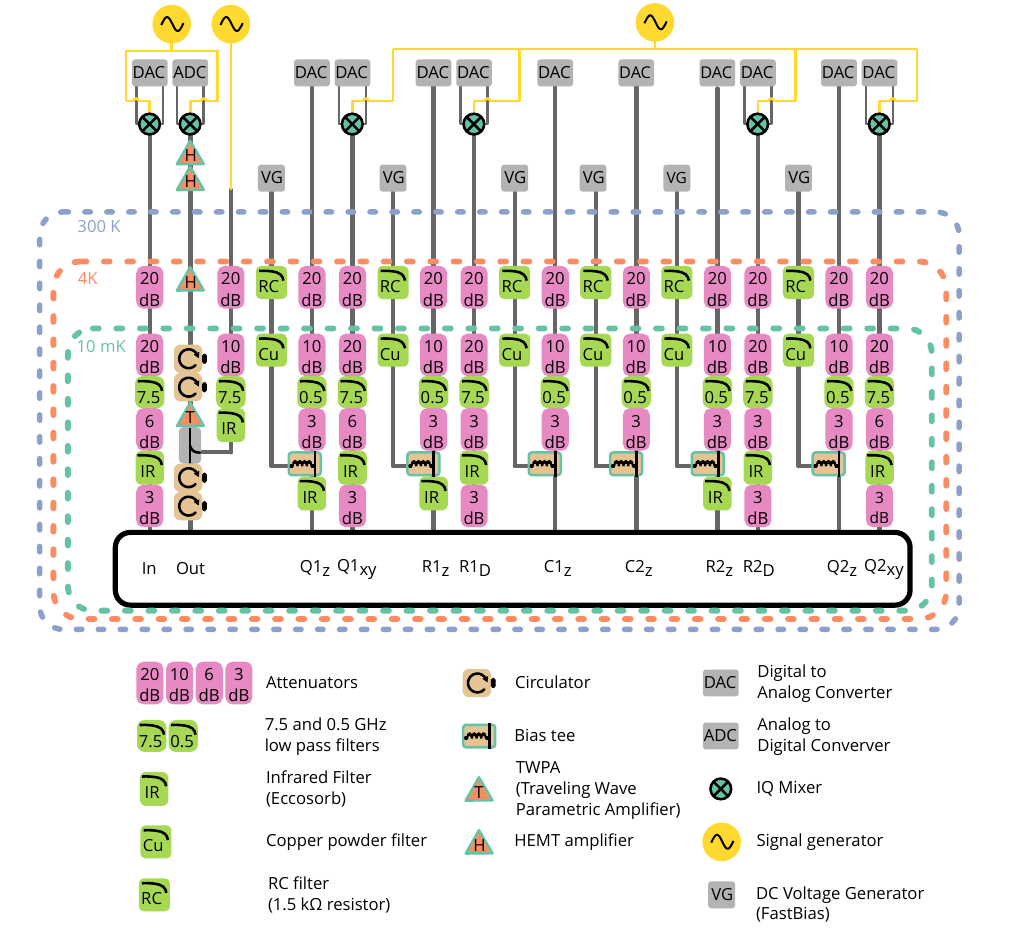}
	\caption{Wiring diagram showing room temperature electronics, attenuators, and amplifiers. Input and output lines are used for qubit state readout. Qubits (Q) and resonators (R) have microwave ($XY$, $D$) lines and current bias ($Z$) lines, with the $Z$ line combined with a low noise continuous bias current component and fast-bias components through a bias tee. Variable couplers only have a current bias line. Microwave signals are generated at room temperature using continuous signal generators that are then modulated by IQ mixers controlled by DACs. Microwave drive lines (In, $XY$, and $D$) are attenuated to reduce thermal noise. Low-noise voltage generators at room temperature are used as current sources by passing signals through a \qty{1.5}{k \ohm} resistor. }
	\label{fig:flyingresonatorwiring}
\end{figure*}

A wiring diagram of the experiment is shown in Fig.~\ref{fig:flyingresonatorwiring}. The qubit readout signals are connected to a TWPA obtained from Lincoln Laboratories with an input noise temperature measured in our lab of around \qty{500}{mK}. Two microwave circulators are placed in front of the TWPA to prevent the qubit from being exposed to the TWPA pump and backwards amplified noise \cite{Esposito2021}.

\section{Device models}
\subsection{Resonator anharmonicity}
Our resonator state preparation, readout and transfer protocols assume low resonator anharmonicity $ \alpha_r $. We estimate $ \alpha_r $ by modeling the nonlinear correction due to a SQUID on one end of a waveguide resonator \cite{Wallquist2006} with a small capacitance to ground on the other end. For a given eigenmode $ n $ of a waveguide resonator of length $ \ell $ we vary the resonator's frequency $ \omega_{n} $  within the range $ (\pi v/\ell) n \le \omega_{n} \le (\pi v/\ell)(n + 1/2)$ by applying a flux through the boundary SQUID which changes its effective inductance. The lower and upper limits correspond to the resonance frequencies for a $ \lambda/2 $ and $ \lambda/4 $ resonator respectively. Here $ v =c/\sqrt{\epsilon_\text{eff}}$ is the transmission line wave velocity, $ c $ is the speed of light in vacuum, and $ \epsilon_\text{eff} $ is the effective dielectric constant for the transmission line. For a coplanar waveguide (CPW) on sapphire in vacuum, $ \epsilon_\text{eff} \approx (1 + \epsilon_r)/2 $ where the sapphire substrate has dielectric constant $ \epsilon_r $. 

Assuming a negligible SQUID capacitance, the resonator's effective capacitance $ C_r $ and inductance $ L_r $ are \cite{Wallquist2006}

\begin{align}
	C_r(x) &= \frac{C_\text{cav}}{2}\left(1 + \frac{\sin(2 x)}{2 x}\right)\\
	L_r(x) &= \frac{L_\text{cav} C_\text{cav}}{x^2 C_r} , 
\end{align}
where $ x = k_{n} \ell $, and $ k_{n} $ is the eigenmode wave vector for the $n$th mode.  The cavity capacitance and inductance are $ C_\text{cav} =  \mathscr{C} \ell $ and $ L_\text{cav} = \mathscr{L} \ell $, respectively, where $ \mathscr{C} = \qty{173}{pF/m} $ and $ \mathscr{L} = \qty{402}{nH/m} $ are the capacitance and inductance per length for a 2:1 center:gap CPW on sapphire \cite{Zhong2019}, respectively. The resonator frequency $ \omega_r $ is 

\begin{equation}
	\omega_r (x) = \frac{x}{\sqrt{C_\text{cav} L_\text{cav}}} .
\end{equation}
The energy deviation $ \delta E_{n} $ of the $ n $th resonator level to first order from the linear resonator is \cite{Wallquist2006}
\begin{align}
	\delta E_{n} (x) &= \frac{-\left(6 n^2 + 6 n + 3\right)}{4} B_k E_C\\
	B_k (x) &= \frac{(1/4)\cos^2(x)}{1 + 2 x/\sin(2 x)} ,
\end{align}
where $ E_C = e^2/(2 C_r) $. The anharmonicity $ \alpha_r $ is then given by the deviation of the first and second levels from the ground and first level, noting that we need to convert energy to frequency via $ \omega = E/\hbar $, as
\begin{equation}\label{tuneres_anharmonicity_eq}
	\hbar \alpha_r =  (\delta E_{2} -  \delta E_{1}) - ( \delta E_{1} -  \delta E_{0}) .
\end{equation}

\begin{figure}
	\centering
	\includegraphics[width=3.37in]{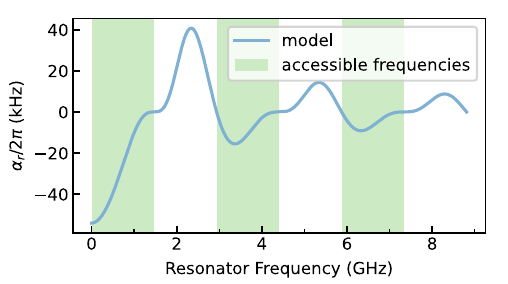}
	\caption{Model for tunable resonator anharmonicity $ \alpha_r/2 \pi $ using Eq.~\ref{tuneres_anharmonicity_eq} for a resonator of fixed length $\ell = $\qty{20.5}{mm} for a 2:1 center:gap CPW on sapphire (see text). The tunable frequency range for the three lowest resonator modes are highlighted in green.}
	\label{fig:figuretuneresmodelanharmonicity}
\end{figure}
We plot the anharmonicity $\alpha_r/2\pi$ in Fig.~\ref{fig:figuretuneresmodelanharmonicity}. According to this model, throughout the full tuning range of the second eigenmode used for the experiments, the magnitude of the anharmonicity does not exceed $ \qty{20}{kHz} $, which is much less than both the resonator/qubit coupling ($\approx \qty{7}{MHz}$) and the resonator/waveguide coupling (maximum $\approx \qty{55}{MHz}$).

\subsection{Classical linear circuit model}

\begin{figure}
	\centering
	\includegraphics[width=3.37in]{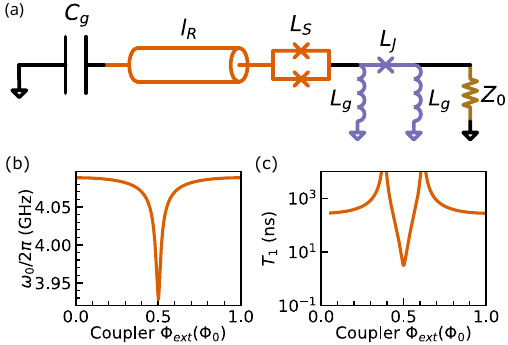}
	\caption{{(a)} Linear circuit model of a tunable resonator coupled to a resistor. A small capacitance to ground $ C_g  = 10^{-14}$ \qty{ }{F} sets the boundary condition for a transmission line resonator (orange) of length $ l_R=\qty{20.5}{mm} $ bounded at the other end by a SQUID with total inductance $ L_S = \qty{0.3}{nH} $. The resonator couples to a $ Z_0 = \qty{50}{\Omega} $ resistor through a variable coupler circuit (green) with inductance to ground $ L_g = \qty{0.2}{nH} $ and variable single junction inductance $ L_J(\Phi_\text{ext}) $. {(b)} Resonant frequency for circuit model in (a) as a function of the external flux through the RF SQUID (variable coupler), see text for inductance to flux expression {(c)} Lifetime for circuit model in (a) as a function of the external flux through the RF SQUID. The minimum lifetime is $ T_1 \approx \qty{3}{ns} $.}
	\label{fig:figurereslinearsimcombined}
\end{figure}

We simulate the system with a simple linear model shown in Fig.~\ref{fig:figurereslinearsimcombined}a to help design parameters that allow us to turn the variable coupling on and off. This linear model is valid for small resonator anharmonicity and for weak coupling \cite{Geller2015} and does not take into account losses in the circuit elements.  The RF SQUID junction inductance is related to the phase across the junction as $L_j = L_{j0}/\cos(\delta)$ with $L_{j0} = \qty{0.6}{nH}$. The junction phase and external RF SQUID flux are related by \cite{Geller2015}

\begin{equation}\label{coupler_flux_eq}
	\delta + \frac{L_l}{L_{T}}\sin(\delta) = 2 \pi \frac{\Phi_\text{ext}}{\Phi_0} . 
\end{equation}

A resistance $ Z_0 = \qty{50}{\Omega} $ is used to model the dissipation load from an infinite transmission line with a characteristic impedance $ Z_0 $ \cite{Pozar2009}. Modeling the circuit as a parallel RLC resonator, we find the circuit resonance $ \omega_p $ numerically when the imaginary part of the total circuit admittance $ Y = 0 $ as shown in Fig.~\ref{fig:figurereslinearsimcombined}b. The effective capacitance $ C_p $ of the resonator at the resonance is given by $ C_p = (1/2)\text{Im}Y'(\omega_p) $ \cite{Nigg2012}. We then calculate the quality factor $ Q_0 = \omega_p Z_0 C_p $ and the resonator lifetime $ T_{1_r} = \omega_r/Q $ in Fig.~\ref{fig:figurereslinearsimcombined}c.

\subsection{Quantum input-output wavepacket model}
To simulate the release of a wavepacket by a resonator we use input-output theory with quantum pulses \cite{Kiilerich2019}. A wavepacket with normalized pulse shape $u(t)$ emitted by the tunable resonator is modeled as a virtual cavity with complex coupling to the tunable resonator 

\begin{equation}\label{key}
	g_u(t) = \frac{u^*(t)}{\sqrt{1 - \int_{0}^{t}dt'\left|u(t')\right|^2}},
\end{equation}
while an incident pulse shape $ v(t)$ couples with

\begin{equation}
	g_v(t) = \frac{-v^*(t)}{\sqrt{\int_{0}^{t}dt'\left|v(t')\right|^2}} .
\end{equation}

We model the wavepacket release and capture as separate processes, in the rotating frame of the (constant) tunable resonator frequency during the process. The total system Hamiltonian is given by \cite{Kiilerich2019}

\begin{equation}
	\hat{H}(t) = \frac{i}{2} \left(\sqrt{\gamma(t)}g^*_u(t)\hat{a}_u^\dagger\hat{c} + \sqrt{\gamma^*(t)}g_v(t)\hat{c}^\dagger\hat{a}_v + g^*_u(t)g_v(t)\hat{a}^\dagger_u\hat{a}_v - \text{h.c.}\right ) , 
\end{equation}
where $\gamma(t)$ is the time dependent resonator decay rate modulated by the variable coupler, $\hat{c}$ is the lowering operator of the tunable resonator and $\hat{a}_u^\dagger = \int dt \ u(t) \hat{a}^\dagger(t)$ is the creation operator for the $u$ virtual cavity with a similar expressions for $v$. 

The Lindblad operator for the capture/emission process is 
\begin{equation}
	\hat{L}_0(t) = \sqrt{\gamma}\hat{c} + g_u(t)\hat{a}_u + g_v(t)\hat{a}_v .
\end{equation}
Additional standard Lindblad operators are used to model the tunable resonator decay and dephasing. The Hamiltonian and Lindlad operators are then used to simulate the system dynamics with a QuTiP master equation solver \cite{Johansson2012}.

\subsection{Wavepacket release and capture}\label{capture_theory}
\begin{figure}
	\centering
	\includegraphics[width=3.37in]{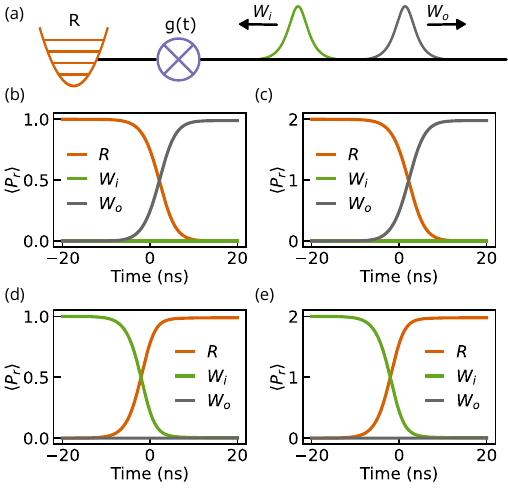}
	\caption{{ (a)} Schematic representing resonator $ R $ coupled to a waveguide through a variable coupling $ g(t) $ with an incoming hyperbolic secant wavepacket $ W_i $ and outgoing wavepacket $ W_o $. { (b,c)} Releasing 1 and 2 photons respectively from $ R $ into $ W_o $ with high efficiency ( 99\% population transfer) using the same release parameters. { (d,e)} Capturing 1 and 2 photons respectively into $ R $ from $ W_i $ with high efficiency ( 99\% population transfer) using the same capture parameters. }
	\label{fig:catchreleasesims}
\end{figure}

Using input-output formalism with quantum pulses \cite{Kiilerich2019} we simulate the release and capture of a single excitation with near unit transfer efficiency using a hyperbolic secant pulse shape, $ u(t)\propto \text{sech}(\kappa_c t/2)$, where $\kappa_c $ is the characteristic bandwidth of the wavepacket (Fig.~\ref{fig:catchreleasesims}b and c). The cavity energy relaxation to the waveguide $ \kappa (t) $ that maximizes emission into this wavepacket is given by \cite{Bienfait2019}
\begin{equation}
	\kappa (t) = \kappa_m \frac{e^{\kappa_c (t - t_0)}}{1 + e^{\kappa_c (t-t_0)}} , 
\end{equation}
where $ \kappa_m $ the maximum coupling strength and $ t_0 $ varies the target output wavepacket in time. $ \kappa (t) $ monotonically increases and then stays constant at its maximum value. We vary $ \kappa_m $ and $ t_0 $ to maximize population into the wavepacket\footnote{If we vary $ t_0 $ away from release window, the resonator would still decay as expected, but there would not be any excitations contained in the time envelope for the wavepacket shape we defined}. In Fig.~\ref{fig:catchreleasesims}, $ 1/\kappa_c =\qty{2}{ns}$ , $ \kappa_m = \qty{0.60}{GHz} $, $ t_0 = \qty{2.62}{ns} $, and have transfer efficiencies $ \approx 0.99 $. The discrepancy from unity efficiency is likely a numerical issue as the expression is analytically derived. Keeping all parameters constant in this numerical simulation except for the number of photons in the incoming pulse we capture the pulse with a similar efficiency. Similarly, it is possible in simulation to capture both single and multi-photon excitations from the sample wavepacket shape without changing experimental parameters after optimizing the single excitation catch. 

\section{Pulse sequences}
\begin{figure}
	\centering
	\includegraphics[width=3.37in]{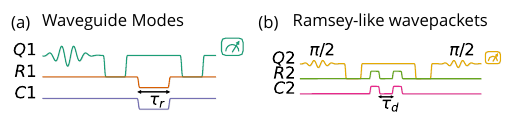}
 \caption{ Pulse sequences for Fig. 2 in the main text with qubits $ Q1 $ and $ Q2 $, resonators $ R1 $ and $ R2 $ and variable couplers $ C1 $ and $ C2 $. Microwave drives are depicted as oscillatory Gaussian pulses. The other shapes are applied fluxes through SQUIDs to tune component frequencies or coupling strength. Measured qubits are labeled with a measurement symbol at the end of sequences.{(a)} Waveguide modes seen from $R1$. After $Q1$ is excited, the excitation is swapped to $R1$, the coupler is turned on with weak coupling to the waveguide while the resonator frequency is shifted for time $\tau_r$, the excitation is swapped back to $Q1$ and $Q1$ is measured. {(b)} Ramsey-like measurement of itinerant wavepackets. $Q2$ is prepared in a $|0\rangle + |1\rangle$ superposition which is swapped to $R2$. The excitation is released as a wavepacket with constant \qty{20}{ns} rectangular pulses turning on the coupling while applying a flux to the resonator SQUID to change its frequency. After swapping back to $Q2$ a final $\pi/2$ pulse is applied before measurement.}
 	\label{fig:fig2pulsesequences}
\end{figure}

\begin{figure}
	\centering
	\includegraphics[width=3.37in]{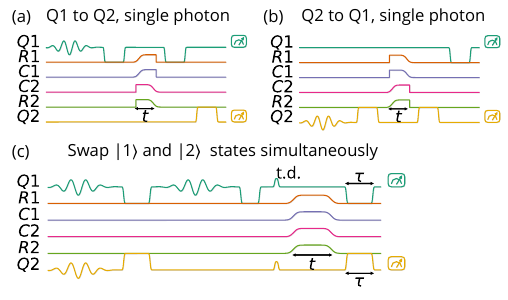}
 \caption{Pulse sequences for Fig. 3 in the main text. {(a)} Single excitation transfer from $Q1$ to $Q2$. After exciting $Q1$, swap the excitation to $R1$, release the itinerant wavepacket from $R1$ and capture with $R2$, and swap population from both resonators to their respective qubits before measurement. {(b)} Similar to (a) in the opposite direction. {(c)} Simultaneous state swap between resonators. Excite both qubits and swap to respective resonators. $Q1$ is excited again and swapped into $R1$. A thermal dump t.d. resets the states of the qubits in preparation for tomography. The states from the both resonators are simultaneously released and after time $t$ caught by the other resonator. Both qubits are tuned into resonance with the resonators for time $\tau$ and the qubit oscillations are fit to extract the resonator population distribution. }
 	\label{fig:fig3pulsesequences}

\end{figure}

\begin{figure}
	\centering
	\includegraphics[width=3.37in]{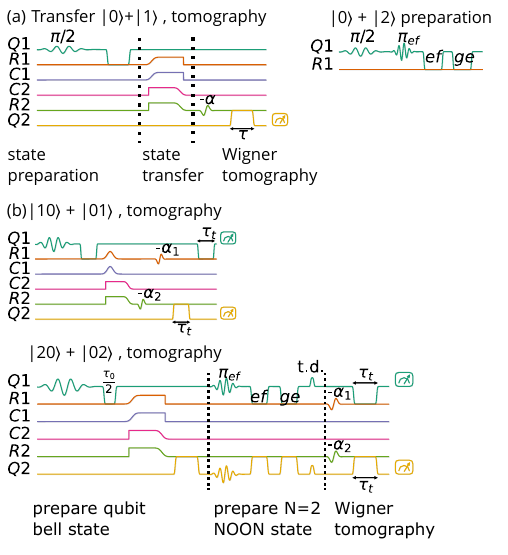}
 \caption{ Pulse sequences for Fig.~4 in the main text. {(a)} Prepare $ |0\rangle + |1\rangle $ superposition state in $ R1 $, transfer the state, then displace $ R2 $ with a coherent pulse with amplitude $ -\alpha $, then interact with qubit for time $ \tau $ for Wigner tomography. The sequence for $ |0\rangle + |2\rangle $ state preparation for $ R2 $ is shown and state transfer and Wigner tomography is otherwise identical to the $ |0\rangle + |1\rangle $ sequence. {(b)} Fig.~3b, Create an entangled resonator Bell state $|10\rangle + |01\rangle$ by swapping a full excitation from $ Q1 $ to $ R1 $, releasing half the resonator state into the waveguide as a shaped wavepacket and capturing at $ R2 $, then measure resonators with joint Wigner tomography. For the $|20\rangle + |02\rangle$ state we first create an entangle qubit Bell state by swapping half of a full excitation from $ Q1 $ to $ R1 $, transferring the resonator state, and then swapping the state from $ R2 $ to $ Q2 $. The $ N=2 $ NOON state is created by driving the $ ef $ qubit transition on both qubits, then swapping the qubit state to the resonator. We finally measure both resonators with joint Wigner tomography. }
 	\label{fig:fig4pulsesequences}
\end{figure}

	Pulse sequences for Fig.~2, 3, and 4 in the main text are shown in Fig.~\ref{fig:fig2pulsesequences},Fig.~\ref{fig:fig3pulsesequences}, and Fig.~\ref{fig:fig4pulsesequences} respectively.  

\section{Device parameters}

The qubit and readout parameters are given in table~\ref{flyingres_qubit_params}. The resonator and coupler parameters are given in table~\ref{flyingres_res_params}. The qubit state is measured via dispersive readout through a single stage asymmetric bandpass Purcell filter with inductively coupled input and output \cite{Sank2014,Yan2023}.

\begin{table*}[ht]
	\begin{center}
		\begin{tabular*}{5.5in}{@{\extracolsep{\fill}} l c c}
			\hline
			\hline
			Qubit parameters & Qubit 1 & Qubit 2\\
			\hline 
			Qubit bare frequency, $\omega_{qb}/2\pi$  &  \qty{6.250}{GHz} &  \qty{6.242}{GHz}\\
			Qubit capacitance, $ C_q $ (design value) &  \qty{90}{fF} &  \qty{90}{fF}\\
			Qubit anharmonicity $ \alpha_q $ & -\qty{192}{MHz} & -\qty{192}{MHz}\\
			Qubit lifetime, $ T_{1q} $  & \qty{20}{\mu s} & \qty{22}{\mu s}\\
			Qubit Ramsey dephasing time, $ T_{2q, \text{Ramsey}} $  & \qty{2.62}{\mu s} & \qty{0.56}{\mu s}\\
			Qubit spin-echo dephasing time, $ T_{2 E} $ (\qty{}{\mu s}) & \qty{4.20}{\mu s} & \qty{3.95}{\mu s}\\
			$ |e\rangle $ state readout fidelity, $ F_e $ & 0.95 & 0.96\\
			$ |g\rangle $ state readout fidelity, $ F_g $ & 0.98 & 0.97\\
			Thermal excited state population & 0.010 & 0.026\\
			Readout resonator frequency, $ \omega_{ro}/2\pi $& \qty{5.358}{GHz} & \qty{5.312}{GHz}\\
			Readout resonator decay rate (design value), $ \kappa_r/2\pi $ & \qty{4.58}{MHz} & \qty{4.58}{MHz}\\
			Readout dispersive shift, $ \chi/2\pi $ & \qty{1.4}{MHz} & \qty{1.3}{MHz}\\
			\hline
			\hline
			\\
			\hline
			\hline
			Purcell filter parameters & value\\
			\hline
			Input quality factor (design value), $ Q_{fi} $ & $ \sim2000 $\\
			Output quality factor (design value), $ Q_{fo} $ & $ 15.5 $\\
			Purcell filter frequency (design value), $ \omega_p/2\pi $ & \qty{5.29}{GHz}\\
			\hline
			\hline
		\end{tabular*}
	\end{center}
	\caption{\label{flyingres_qubit_params} { Qubit and readout parameters. } }
\end{table*}

\begin{table*}[ht]
	\begin{center}
		\begin{tabular*}{5.5in}{@{\extracolsep{\fill}}l c c}
			\hline
			\hline
			Variable resonator parameters & Resonator 1 & Resonator 2\\
			\hline
			Qubit/resonator coupling, $ g_{qr}/2\pi $ & \qty{6.805}{MHz} & \qty{6.830}{MHz}\\
			Resonator lifetime (\qty{4.058}{GHz}), $ T_{1r} $ & \qty{4.57}{\mu s}& \qty{0.86}{\mu s}\\
			Resonator Ramsey dephasing time (\qty{4.058}{GHz}), $ T_{2r, \text{Ramsey}} $ & \qty{0.95}{\mu s} & \qty{0.90}{\mu s}\\
			Resonator bare frequency $ \omega_r/2\pi $ & \qty{4.269}{GHz} & \qty{4.269}{GHz}\\
			Resonator DC SQUID inductance (design value), $ L_{rs} $ & \qty{0.25}{nH} & \qty{0.25}{nH}\\
			\hline
			\hline
			\\
			\hline
			\hline
			Variable coupler parameters & Coupler 1 & Coupler 2\\
			\hline
			Coupler junction inductance (design value) $ L_{cj} $ & \qty{0.6}{\nano H} & \qty{0.6}{\nano H}\\
			Coupler grounding inductance, $ L_g $ (design value) & \qty{0.2}{\nano H} & \qty{0.2}{\nano H}\\
			Coupler stray inductance, $ L_g $ (assumed design value) & \qty{0.1}{\nano H} & \qty{0.1}{\nano H}\\
			\hline
			\hline
		\end{tabular*}
	\end{center}
	\caption{\label{flyingres_res_params} { Resonator and coupler parameters. } }
\end{table*}

\subsection{Resonator Control}

\begin{figure}
	\centering
	\includegraphics[width=3.38in]{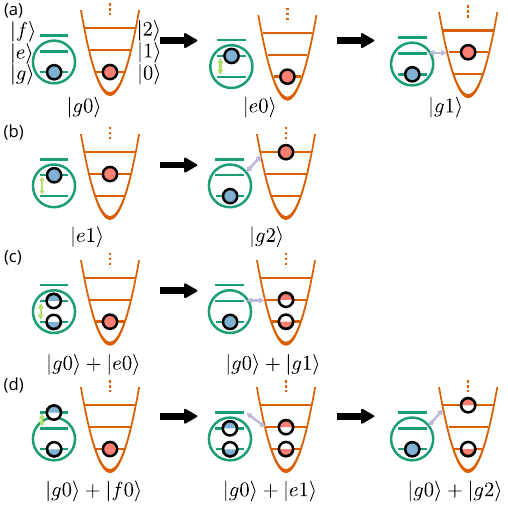}
	\caption{{ Resonator Fock and superposition state preparation}{ (a)} Steps to prepare a single photon state by swapping excitations from a transmon qubit. The qubit is depicted on the left (blue) and resonator on the right (red). Both systems are initially prepared in their respective ground states. The qubit is excited to $ |e\rangle $, and then that excitation is swapped into the resonator in time $ \tau_\text{swap} $ with both systems on resonance. { (b)} A two-photon state in the resonator is prepared by starting with the single photon state in (a), exciting the qubit, then swapping the excitation to the resonator in time $ \tau_\text{swap}/\sqrt{2} $. { (c)} A resonator $ |0\rangle + |1\rangle $ state is prepared by first preparing $ |g\rangle + |e\rangle $ in the qubit, then swapping the excited state to the resonator. { (d)} A resonator $ |0\rangle + |2\rangle $ is prepared by starting from the qubit $ |g\rangle + |e\rangle $ state (first step in (c)) then exciting the qubit to $ g\rangle + |f\rangle $. Two swaps into the resonator (both  $ \tau_\text{swap}/\sqrt{2}$) then creates a $ |0\rangle+ |2\rangle $ resonator state. }
	\label{fig:figureresonatorstateprepcartoon}
\end{figure}

We prepare resonator states by resonantly swapping excitations using the frequency-tunable qubit \cite{Liu2004,Hofheinz2008,Hofheinz2009}. With the resonator initially in the ground state, a qubit excitation transfers to the resonator in time $ \tau_0 = \pi/2g_{qr} $, with $g_{qr}/2\pi = \qty{6.8}{MHz}$ the qubit/resonator coupling strength (Fig.~\ref{fig:figureresonatorstateprepcartoon}a). For a resonator initially prepared in a higher excited state $ n $ this transfer time becomes shorter as $ \tau_n = \tau_0/\sqrt{n} $. By repeatedly exciting and then swapping excitations from the qubit to the resonator, multi-photon resonator states are prepared. (Fig.~\ref{fig:figureresonatorstateprepcartoon}b). 

Fig.~\ref{fig:figureresonatorstateprepcartoon}c-d further show how $ |0\rangle + |N\rangle $ resonator superposition states are prepared using an ancilla qubit \cite{Wang2011}. A  $ |0\rangle + |1\rangle $ resonator state is made by resonantly interacting the qubit in state $ |g\rangle + |e\rangle $ with the resonator initially in the ground state for time $\tau_0$. To prepare larger $ |0\rangle + |N\rangle $ resonator states we first excite the transmon from $ |g\rangle + |e\rangle $ to $ |g\rangle + |f\rangle $, and then swap the highest qubit excitation into the resonator. If $N=2$, we swap the remaining qubit excitation into the resonator (Fig.~\ref{fig:figureresonatorstateprepcartoon}d). Otherwise, for higher $N$ we excite the $ |e\rangle -|f\rangle $ transition and swap to the resonator for the additional excitations. 

\section{Device characterization}

\subsection{Qubit/Resonator coupling}
\begin{figure}
	\centering
	\includegraphics[width=3.38in]{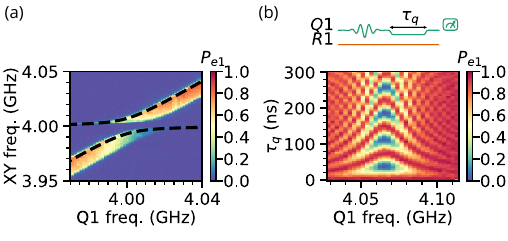}
	\caption{{ (a)} Q1 two-tone spectroscopy showing avoided level crossing with R1. The peak is broadened, likely due to pulse distortion in the qubit flux bias. We fit the avoided level crossing by eye to be approximately $ g/2\pi \approx \qty{7}{MHz} $. { (c)} After correcting for pulse distortion \cite{Chen2018}, the qubit/resonator swaps show the characteristic chevron shape with $g/2\pi = \qty{6.8}{MHz}$.}
	\label{fig:figureresonatortuneup}
\end{figure}

Fig.~\ref{fig:figureresonatortuneup}a shows the $Q1/R1$ two-tone spectroscopy \cite{Wallraff2007} avoided level crossing with coupling strength \qty{6.8}{MHz}.

\subsection{Resonator characterization}
\begin{figure}
	\centering
	\includegraphics[width=3.37in]{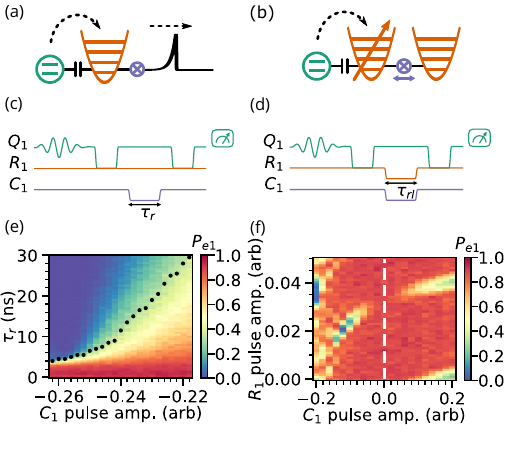}
	\caption{Cartoon {(a)}, pulse sequence {(c)}, and data {(e)} of a single excitation decaying from $R1$ into the waveguide. $R1$ is coupled to the waveguide with varied coupling strength controlled by $C1$, for time $\tau_r < \qty{30}{ns}$ before reflections returning from the far end of the waveguide are visible. Black dots show when $R1$ population is $1/e$ the value at $\qty{0}{ns}$, showing that population is released into the waveguide with characteristic time under $\qty{4}{ns}$. Cartoon {(b)}, pulse sequence {(d)}, and data {(f)} for method to find zero coupling point between the resonator and waveguide, shown as a dotted line in {(f)}. After swapping a single excitation into the resonator we set the delay time $ \tau_{rl} = \qty{200}{ns} $ and then vary the $C1$ coupling strength until the $R1$ population is independent of frequency.}
	\label{fig:couplertuneup}
\end{figure}

\begin{figure}
	\centering
	\includegraphics[width=3.37in]{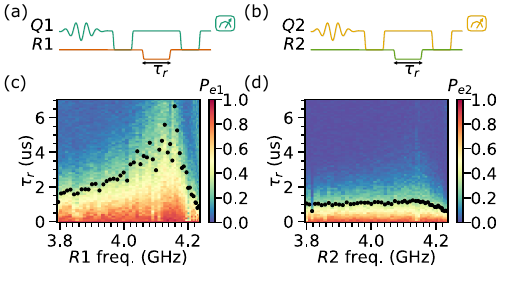}
	\caption{Pulse sequence {(a)} and data {(b)} for $ T_1 $ measurement of $ R1 $ as a function of its frequency. Black dots are fits to exponential decay. A sharp decrease in lifetime above \qty{4.2}{GHz} possibly a box or chip mode interacting with the resonator. {(c)} $ T_1 $ measurement of $ R2 $. $ T_1 $ is flat with frequency below \qty{4.2}{GHz} (with a similar drop as in (a) above \qty{4.2}{GHz} and similar shape at long delay $ \tau_r $). This may indicate that the resonator's DC SQUID is not the dominant loss channel.}
	\label{fig:reslifetimes}
\end{figure}

In Fig.~\ref{fig:couplertuneup} $R1$ is tuned to both maximum and minimum coupling to the waveguide, demonstrating a wide range of control for the wavepacket emission. With the coupling to the waveguide off, Fig.~\ref{fig:reslifetimes} shows both resonator lifetimes as a function of the resonator frequencies. Around \qty{4}{GHz}, both resonator lifetimes are above \qty{1}{\mu s} which is suitable for our experiments. Above \qty{4.2}{GHz} is a sharp decrease in the resonator lifetimes, possibly a box mode due to the large chip package. The reduction in the $T_1$ for $R1$ has been seen with similar tunable resonator designs\cite{Wang2013}.

\subsection{Waveguide Characterization}\label{waveguide_char}
\begin{figure}
	\centering
	\includegraphics[width=3.38in]{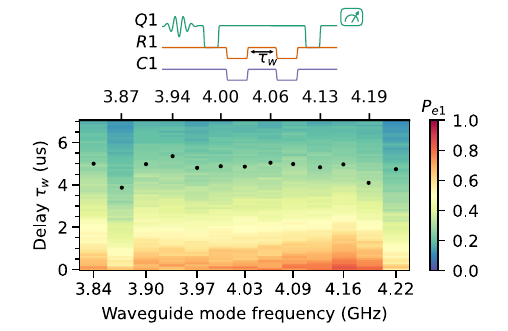}
	\caption{Waveguide mode $ T_1 $ measurement. A one photon excitation from $ Q1 $ is swapped first into $ R1 $ and then into a waveguide mode. After waiting variable time $ \tau_w $, the excitation is swapped back to $ R1 $ and then to $ Q1 $.}
	\label{fig:figuretlinelifetime}
\end{figure}

We swap single excitations into single waveguide standing modes with frequencies across several hundred \qty{}{MHz} to characterize their loss as shown in Fig.~\ref{fig:figuretlinelifetime}. The standing mode loss measurements strongly suggests that waveguide loss is not the dominant source of infidelity in the experiment.   

\subsection{Wavepacket Characterization} \label{wavepacket_char}
\begin{figure}
	\centering
	\includegraphics[width=3.38in]{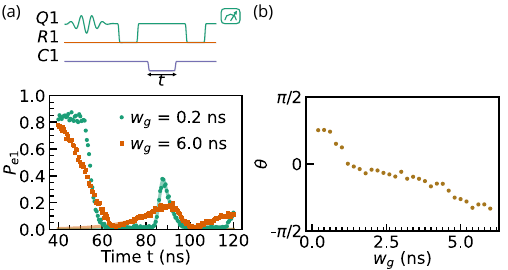}
	\caption{{ (a)} Releasing a wavepacket with a flattop pulse of varying Gaussian rise widths $ w_g $. The wavepacket bounces off the other end of the waveguide and we measure the resonator population with the coupler fully on. The wavepackets are then fit to a skewed hyperbolic secant with skew $ \theta $ shown in (b), excluding the data points from release and second reflection.{ (b)} Varying $ w_g $ and fitting the wavepackets we see a decreasing trend between $ w_g $ and $ \theta $.}
	\label{fig:figurewavepackets}
\end{figure}

Before optimizing the release and capture of the wavepackets, we roughly estimate the wavepacket shape by applying a flattop pulse (Gaussian convolved with rectangular pulse) with varying rise times $ w $ and then monitor the resonator population as a function of time while strongly coupled to the waveguide, as shown in Fig.~\ref{fig:figurewavepackets}a. This is not in general an optimized coupling shape, however it gives us a sense of the range of wavepacket shapes we can generate. If the coupling to the waveguide is ramped up quickly, then the release is dominated by an exponential decay and lead to an asymmetric curve. For slow Gaussian ramp up the asymmetry is in the opposite direction. We characterize this asymmetry by modeling the wavepacket as a skewed hyperbolic secant $ f(t,\theta,w) $ with skew $ \theta $ and width $ w $. 

\begin{equation}\label{skew_eq}
	f(t) = \cos \theta \frac{e^{\frac{\theta (t - t_0)}{w}}}{2 \cosh \left(\frac{\pi (t - t_0)}{2w}\right)} . 
\end{equation}
The exponential adds a skew to a hyperbolic secant function, and the cosine normalizes the integral of the function to one. This function is not derived from the experimental circuit, it simply happens to fit the data reasonably well and lets us extract the skew of the wavepacket from a fit. Ideally, the skew should be zero for a time-symmetric wavepacket. Fitting the wavepackets with this function as shown in Fig.~\ref{fig:figurewavepackets}b, we find wavepackets with varying asymmetries. This is only a rough estimate for the wavepacket shape as the time to turn off the coupling is finite and so distorts the measured pulse shape.

\section{State transfer}
\subsection{Single excitation transfer, Swaps}
We also transfer energy between resonators by swapping into a single waveguide mode. After exciting qubit 1, swapping into resonator 1,  a single waveguide mode at \qty{4.157}{GHz}, resonator 2 and finally qubit 2, we measure around 0.78 population in the second qubit. Numerical simulations of this transfer that include a 94\% visibility,and qubit/resonator decay and dephasing predict a 0.84 population in qubit 2 for the same process. The rest of the discrepancy may be due to incomplete energy transfer between components.

\subsection{Single excitation transfer, wavepacket}
 We optimize the release and capture pulses for single excitations by maximizing population transfer, and then use the same parameters for general resonator state transfer. Gaussian filters in our wiring smooth arbitrarily generated pulses on the order of \qty{3}{ns} limiting the time resolution for pulse control and making it difficult to reproduce analytic pulse shapes and timing for optimal state transfer as in Ref.~\cite{Bienfait2019}.  Here we use a Bayesian optimization package to vary the pulse shape in the time domain for release and capture \cite{Nogueira2014}, maximizing the total transferred amplitude from an initial single excitation in the releasing resonator. This optimization for wavepacket release and capture is done independently for both transfer directions.

\subsection{Wigner tomography}\label{wigner_experiment}
The Wigner function $ W(\alpha) $ is a quasiprobability distribution for a resonator's quantum state which can be calculated from the density matrix $ \rho $ \cite{Haroche2006}

\begin{equation}\label{wigner_eq}
	W(\alpha) = \frac{2}{\pi} \text{Tr}\left[D(-\alpha) \rho D(\alpha) \mathcal{P}\right] , 
\end{equation}
where $ \text{Tr} $ is the trace, $ D(-\alpha) = D^\dagger (\alpha) = \exp (\alpha^* a - \alpha a^\dagger) $ is the displacement operator for the resonator, $ \mathcal{P} $ is the resonator's photon number parity, and $ \alpha = x + ip $ is the complex phase space coordinate.  $ x $ and $ p $ are the resonator's position and momentum variables respectively. Applying the displacement operator on the resonator ground state $ |0\rangle $ creates coherent states $ |\alpha\rangle $.  Experimentally, $ \alpha $ is also the coherent displacement of the resonator with $ -\alpha = (\frac{1}{2})\int \Omega_r (t) dt $, where $ \Omega_r $ is a complex drive amplitude \cite{Hofheinz2009}.

\begin{figure}
	\centering
	\includegraphics[width=3.38in]{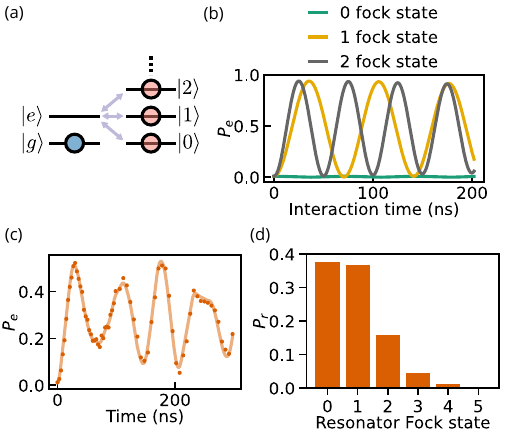}
	\caption{{ (a)} A resonator with unknown state couples to a qubit prepared in the ground state { (b)} Simulation for qubit/resonator oscillations with the resonator prepared in single Fock states. The qubit oscillates  $ \tau_0/\sqrt{n} $, where $ \tau_0 $ is the period for a single excitation { (c)} For an arbitrary resonator state, the qubit population over time is the linear combination of the oscillations that would occur for each prepared Fock state as in (b). Data (points) and fit to model (line) are shown { (c)} Resonator occupation probability distribution from fitting the data in (c). We see that the resonator has a Fock distribution consistent with a coherent state with $ \alpha \approx 1 $.}
	\label{fig:figurewignertomographysteps}
\end{figure}

We displace the resonator by capacitively coupling a signal line to the resonator. The Fock state probability distribution is measured by resonantly interacting the qubit with the resonator and fitting the oscillations to a system model\cite{Hofheinz2009,Wang2011,Liu2004}, as shown in Fig.~\ref{fig:figurewignertomographysteps}. The resonator parity is then calculated from that distribution. 

If the resonator is prepared in a well-defined Fock state, then a qubit prepared in the ground state goes through Rabi oscillations with a period of $ \tau_0/\sqrt{n} $, where $ \tau_0 $ is the period for a single excitation. If the resonator is prepared with a superposition of Fock states, the qubit oscillates following a linear combination of these oscillations proportional to their associated resonator populations. 

To reconstruct the density matrix from the Wigner function we use a MATLAB package \cite{Grant2014} that uses convex optimization to find the most likely $ \rho $ given the measured photon number distributions as a function of displacement \cite{Strandberg2022}. During this reconstruction procedure we constrain $ \rho $ to be Hermitian, positive semi-definite, and have a trace equal to 1.
The joint resonator density matrix reconstruction for NOON states further constrain elements with photon indices larger than N to zero \cite{Wang2011}.

\begin{figure}
	\centering
	\includegraphics[width=3.38in]{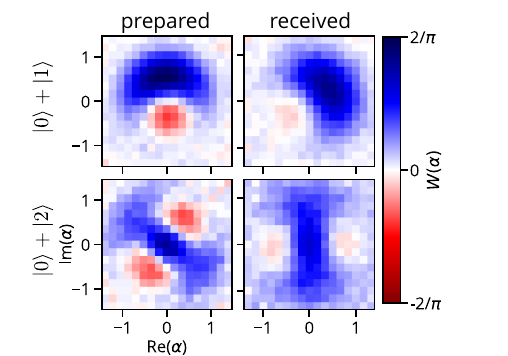}
	\caption{Wigner tomograms using resonator parity data for Fig.~2a in the main text.}
	\label{fig:transfersuperpositionsparity}
\end{figure}

\subsection{Microwave displacement calibration and crosstalk}
\begin{figure}
	\centering
	\includegraphics[width=3.37in]{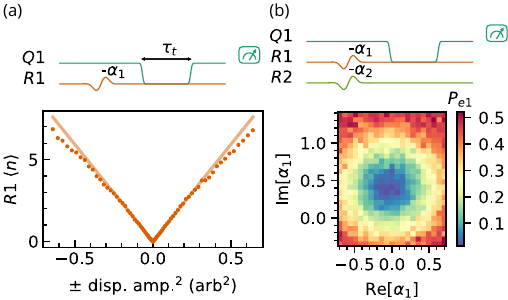}
	\caption{{ (a)} Calibration measurement for $ \alpha $. We displace the resonator and measure the average population of the resonator. Fitting the time trace with a qubit/resonator model returns a linear relationship at low photon powers with the square of the displacement amplitude. The linear fit here is up to $ 0.4 $ displacement amplitude which roughly corresponds to $ n \approx 1.9 $. { (b)} Microwave crosstalk node 2 on node 1. We apply a constant displacement pulse $ \min\alpha_2 $ with $ R2 $'s displacement line that would correspond to a photon population in $ R2 $ of $ \langle n_2 \rangle \approx 2.7 $. At the same time we apply a displacement $ \alpha_1 $ on resonator 1, and then swap the resulting population into $ Q1 $ at the $ 1 $ photon swap time $ \tau_0 = \qty{38}{ns}$. When $ \alpha_1 $ cancels the crosstalk from the displacement from node 2 we measure 0 population in $ Q1 $.}
	\label{fig:figureresonatordisplacements}
\end{figure}

Using the resonant photon state reconstruction technique we calibrate the resonator displacement pulses. Fig.~\ref{fig:figureresonatordisplacements}a shows the average photon number after applying a \qty{20}{ns} displacement pulse and fitting the qubit/resonator interaction. Displacement amplitude $ \propto V $ (proportional to $ V $), and since the photon number $ n \propto V^2 $, $ n \propto $ the square of the displacement amplitude. At higher photon numbers the linear relationship deviates, possibly due to either resonator nonlinearity or photon number fitting errors. For a coherent state $ \langle n_\alpha \rangle = |\alpha|^2 $, so the slope of this line gives the calibration factor between $ \alpha $ and the displacement amplitude, in the case of Fig.~\ref{fig:figureresonatordisplacements}a, the slope is $ 3.44 $.

Microwave crosstalk occurs when we want to only apply a microwave tone to one component but affect other components as well. This may occur because of unintended coupling in either the microwave lines or sample packaging.  In our system the two qubits are detuned from each other by tens of \qty{}{MHz} so that microwave crosstalk between the two is minimal. This is confirmed by separately exciting the two qubits and jointly measuring them to monitor the ground state populations. The tunable resonators are detuned from the qubits by several hundred \qty{}{MHz} when applying microwave pulses, and so have negligible microwave crosstalk between the two. However, the tunable resonators are at the same frequency, and so there is non-negligible crosstalk between them.  When we apply a microwave pulse to displace resonator 1 with amplitude $ a_1 $ and phase $ \theta_1 $, we also displace resonator 2 with amplitude $ a_{12} $ and phase $ \theta_{12} $. Similarly there is crosstalk in the other direction. We correct for this crosstalk when we need to simultaneously measure the state of both resonators, as in the NOON state experiments.

To measure this crosstalk we use a similar procedure that we use for calibrating displacement amplitude to resonator $ \alpha $ for a single resonator: Wigner tomography with the resonator prepared in the ground state, and scanning the displacement amplitude. However, instead of displacing the resonator using its own displacement line, we use the microwave crosstalk from the other resonator displacement line. We measure the phase of this crosstalk relative to the 0 phase for each resonator, by applying displacements from both displacement lines and measuring at which phase the two tones cancel each other as in Fig.~\ref{fig:figureresonatordisplacements}b. This crosstalk is quite large, when applying a displacement pulse -$ \alpha_2 $ to populate resonator 2 with \qty{1}{photon} we also populate resonator 1 with 0.07 photons. We correct for microwave crosstalk when processing the data.

Flux crosstalk is corrected for qubits and couplers by applying corrective pulses in the experiment.

\bibliography{references_supplement}